\documentclass{article}

\usepackage{arxiv}

\usepackage[utf8]{inputenc} 
\usepackage[T1]{fontenc}    
\usepackage{hyperref}       
\usepackage{url}            
\usepackage{booktabs}       
\usepackage{amsfonts}       
\usepackage{nicefrac}       
\usepackage{microtype}      
\usepackage{lipsum}		
\usepackage{graphicx}
\usepackage{natbib}
\usepackage{doi}
\usepackage{amsmath}

\title{Polytropic Wind Solutions via the Complex Plane Strategy}

\author{{Vasileios~Karageorgopoulos} \thanks{Corresponding author} \\
    Department of Physics,\\
	University of Patras,\\
	Patras, Greece\\
	\texttt{vkarageo@upatras.gr} \\
	 \AND
	{Konstantinos N.~Gourgouliatos} \\
    Department of Physics,\\
	University of Patras,\\
	Patras, Greece\\
	\texttt{kngourg@upatras.gr} \\
	 \And
	{Vassilis Geroyannis} \\
    Department of Physics,\\
	University of Patras,\\
	Patras, Greece\\
	\texttt{vgeroyan@upatras.gr} \\
}

\date{}

\renewcommand{\headeright}{}
\renewcommand{\undertitle}{}
\renewcommand{\shorttitle}{Polytropic Wind Solutions via the Complex Plane Strategy}

\begin{document}
\maketitle

\begin{abstract}
Solar--type stars generate spherical winds, which are pressure driven flows, that start subsonic, reach the sound speed at the sonic point and transition to supersonic flows. The sonic point, mathematically corresponds to a singularity of the system of differential equations describing the flow. In the problem of an isothermal wind, the Parker solution provides an exact analytical expression tuned appropriately so that the singularity does not affect the solution. However, if the wind is polytropic it is not possible to find an analytical solution and a numerical approach needs to be followed.
We study solutions of spherical winds that are driven by pressure within a gravitational field. The solutions pass smoothly from the critical point and allow us to study the impact of the changes of the polytropic index to these winds. We explore the properties of these solutions as a function of the polytropic index and the boundary conditions used.
 We apply the Complex Plane Strategy (CPS) and we obtain numerically solutions of polytropic winds. This allows us to avoid the singularity appearing in the equations through the introduction complex variables and integration on the complex plane.
 Applying this method, we obtain solutions with physical behaviour at the stellar surface, the sonic point and at large distances from the star. We further explore the role of the polytropic index in the flow and the effect of mass--loss rate and temperature on the solution. We find that the increase of polytropic index as well as the decrease of flow parameter both yield to a smoother velocity profile and lower velocities and shifts the transition point from subsonic to supersonic behaviour further from the star. Finally, we verify that the increasing of coronal temperature yields higher wind velocities and a weaker dependence on polytropic index.

\end{abstract}


\section{Introduction}
\label{intro}

The seminal work of \cite{P1958} has set the foundations for the study of the solar wind and provided an analytical solution to the problem of a spherically symmetric, isothermal, pressure--driven wind. The wind starts as a subsonic flow from the surface of the Sun, gets accelerated due to the thermal pressure gradient, exceeds the sound speed and becomes a supersonic flow. While our current understanding of the the solar is enriched by the observations accumulated over half a century \citep{Kohl:1998, Marsch:2006}, Parker's picture of a pressure driven wind encapsulates a substantial part of the underlying physics. This picture has been extended by considering a polytropic gas equation of state (EOS), rotation
and magnetic field treated with analytical and numerical solutions \citep{P1960, P1964a, P1964b,  P1964c, P1965, P1966, Mel2004} and numerical solutions, e.g.~\cite{Weber1967, Suess:1975, Keppens1999, K2011, Wat2012}. 

While the main motivation and application of these studies has been the solar wind, stellar winds have been explored in several types of stars with radiation--driven winds \citep{Castor:1975, Mattsson:2010}. For instance, in the CAK formalism \citep{Castor:1975} the Parker analysis is enriched with radiation pressure. In more recent attempts, non--local--thermal equilibrium effects are included \citep{Sundqvist:2019}.
Such winds impact the stellar environment including extra--solar planets and their atmospheres \citep{Preusse:2005, See:2014} but also the winds originating from extra--solar planets themselves \citep{Tian2005, Stone2009}. All these models explore a wide variety of parameters, beyond the ones measured in the solar system, creating an array of models applicable to diverse physical conditions. Furthermore, the problem of spherical accretion formulated by Bondi \cite{Bondi:1952} is mathematically similar to the wind problem except for the boundary conditions that correspond to zero velocity at infinity and maximum supersonic velocity at the surface of the star. Among the variety of wind prescriptions, here, we restrict our simulations to thermally driven winds to present the CPS method as a mathematical tool that can be implemented in this area of research.

From a mathematical perspective, the sonic point corresponds to a singularity of the equations. In the case of an isothermal wind, the integration of the equations through the singularity is straight-forward, leading to the Bernoulli integral and a smooth solution. The introduction of a polytropic index, other than $\gamma=1$ that corresponds to the isothermal wind, leads to a more complicated problem that can only be solved numerically, or approximately through a semi--analytical approach. Numerical solutions are not straightforward due to the presence of the critical point. For instance, application of the shooting method and integration forward from the stellar surface requires fine tuning to converge to the solution that passes smoothly through the critical point. The numerical solution of \cite{Keppens1999} employs a different approach by integrating the dynamical evolution of the time-dependent Euler equations until the system relaxes to an equilibrium which then corresponds to the solar wind solution. While this approach leads to solutions that pass smoothly from the sonic point, it is computationally more demanding, being time--dependent,  compared to a solution of the equilibrium differential equations. An other option is the integration on either side of the critical point and interpolation of the two solutions so that they match at the critical point \citep{P1964a}. This approach provides a smooth curve, however, it does not address the equation as the critical point and the solution found may correspond to a different curve, which is similar to the solution in question, but with some deviations at the critical point. 

In this paper, we propose and implement an alternative approach by integrating the equation on a complex contour and applying the numerical method of ``complex plane strategy'', (CPS). This method has been used in several other astrophysical initial value problems that suffer from mathematical singularities such as stellar polytropic models \citep{G1988}, solar system and the jovian satellites \citep{G1993, GV1994}), white dwarf models \citep{GH1992}, and the general--relativistic polytropic neutron star models \citep{PG2003, GKa2008, GK2014, GK2015}. The main strength of this method is that it can solve numerically equations with singularities without the need of a piece--wise integration, leading to the omission of the critical points, or the need of fine--tuning, but it rather allows the direct numerical integration of the equation in the entire domain. This comes at the expense of converting the real--valued problem to a complex--valued one and integrating along a contour on the imaginary plane. In this work, we apply the CPS on the problem of a polytropic wind. We reproduce the analytical isothermal solution, using it as a benchmark for the numerical calculation. We extend our study to the polytropic solution and we discuss the results.

The structure of this paper is as follows: We present the mathematical setup of the problem in \S~\ref{setup}, where we also analyse the critical points and illustrate how the CPS treats them. We present our results in section \S~\ref{results} by comparing them against the existing results and extending them to various problems.  We discuss our results in \S~\ref{discussion}. We conclude in section \S~\ref{conclusion}.

\section{Problem setup}
\label{setup}

\subsection{Isothermal wind}
\label{isosetup}

The simplest model for the solar wind is that of an isothermal spherically symmetric flow described by a system of two ordinary differential equations arising from the conservation laws of mass and momentum respectively. The equations take the following form (\cite{P1958}, cf. equations~(10), (11))
\begin{equation}
\frac{d}{dr} \left( \rho \, v \, r^2 \right) = 0 \, ,
\label{syseq1}
\end{equation}
\begin{equation}
\rho \, v \frac{dv}{dr} \, + \frac{dP}{dr} \, +\rho \, \frac{G \, M_{\odot}}{r^2} = 0 \, .
\label{syseq2}
\end{equation}
where $P$ is the pressure, $\rho$ is the density, $v$ is the radial velocity, $G$ is Newton's constant, $r$ is the distance from the centre of the Sun and $M_{\odot}$ is the solar mass. As the flow is isothermal, the pressure and density are connected through the following relation $P = \rho v_{s,b}^2$, where the sound speed is expressed in the following form
\begin{equation}
v^2_{s,b}=\left( 2 \, k \, T_b/m_p \right) \, ,
\label{v_so1}
\end{equation}
with $k$ the Boltzmann constant and $m_p$ the proton mass, with protons obeying a Maxwell--Boltzmann distribution of temperature $T_b$. As the flow is isothermal, the sound speed is constant throughout the flow.  

We can conveniently rewrite the system of equations~(\ref{syseq1})--(\ref{syseq2}) in dimensionless form by introducing the following variables; the dimensionless radius $\xi=r/R_{\odot}$ where $R_{\odot}$ is the solar radius, the Mach number $M(\xi)=v(\xi)/v_{s,b}$ and the dimensionless density $\bar{\rho}(\xi)=\rho(\xi)/\rho_b$ where $\rho_b$ is the density of the solar wind at $R_{\odot}$. Substituting the dimensionless quantities, the system take the following form 
\begin{equation}
\frac{d}{d\xi} \left( \bar{\rho} \, M \, \xi^2 \right) = 0 \, ,
\label{syseq3}
\end{equation}
\begin{equation}
\bar{\rho} \, M \frac{dM}{d\xi}\, + \frac{d\bar{\rho}}{d\xi} \, +\lambda\frac{\bar{\rho}}{\xi^2} = 0\, .
\label{syseq4}
\end{equation}

The parameter $\lambda$ is defined as follows \citep{P1958}
\begin{equation}
\lambda = \frac{1}{2}\left( \frac{v_{esc}}{v_{s,b}} \right)^2 = \frac{G \, M_{\odot} \, m_p}{2 \, R_{\odot} \, k \, T_b} \, .
\label{eqlam}
\end{equation}
where the solar escape velocity is $v_{esc}= \sqrt{\frac{2\, G\, M_{\odot}}{R_{\odot}}}$. Indicatively, at the solar corona the above quantities take the values of $v_{esc} \approx 617$km/sec, $v_{s,b} \approx 130$km/sec and $T_b \approx 10^6$K \citep{P1958}. 

Substitution of the continuity equation into the momentum equation leads to an ordinary differential equation for the Mach number as a function of the dimensionless radius $\xi$ 
\begin{equation}
\frac{1}{M^2} \, \frac{dM^2}{d\xi} = \frac{2}{\xi^2} \, \frac{2 \, \xi - \lambda}{M^2-1} \, ,
\label{macheq}
\end{equation}
The above equation is the so--called Mach equation, which can be conveniently written in the following form 
\begin{equation}
\frac{dM}{d\xi} = \frac{M}{\xi^2} \, \frac{2 \, \xi - \lambda}{M^2-1} \, .
\label{veleq}
\end{equation}
It is evident from the above equations that the critical point is located at $r_c=\frac{\lambda}{2}$ where the kinetic energy of an elementary volume is equal its thermal energy. At the critical point the flow velocity becomes equal to sound speed, thus the flow becomes transonic. Beyond this point the solar wind becomes supersonic.

The Mach equation is a non--linear ordinary differential equation that can be integrated by separation of variables. Integration leads to the Bernoulli equation
\begin{equation}
\textrm{ln} \, M -\frac{M^2}{2} = \textrm{ln} \, B - \textrm{ln} \, \xi^2 -\frac{\lambda}{\xi} \, ,
\label{bereq}
\end{equation}
where $B$ is the Bernoulli integral, representing energy. The value of $B$ can be determined by solving equation~(\ref{bereq}) for $M=1$ and $\xi=\lambda/2$, obtaining
\begin{equation}
B_c = \frac{\lambda^2 \, e^{\frac{3}{2}}}{4} \, .
\label{bcrit}
\end{equation}
If we choose a different value for $B\neq B_c$, this will lead us to another family of solutions, either the ``breeze'' solutions that never become supersonic or solutions that remain supersonic for the entire flow. One can also obtain expressions for $f(M,\xi)$ that satisfy equation (\ref{bereq}), but are not physically acceptable as have two different values for the velocity at the same point. Each solution arises from a different set of boundary conditions at $\xi_c$ \citep{Fitz2014}. In particular, if the Mach number at $\xi=\xi_c$ is $M_c > 1$, the corresponding value of solar wind velocity is supersonic everywhere, otherwise, if $M_c < 1$, it remains subsonic everywhere leading to the breeze solutions. We can further obtain the Bondi accretion solution where the velocity is zero at infinity, reaches the sound speed at the critical point and becomes supersonic close to the star, which corresponds to accretion, instead of wind \citep{Bondi:1952}. By applying our method we can discriminate between the wind and Bondi solution by using the Mach number at the critical point as a parameter. We discuss on this issue in \S~\ref{Isores}.

\subsection{Polytropic wind}
\label{polysetup}
The isothermal model postulates that the wind is heated very efficiently and reaches a constant temperature everywhere immediately. If a more realistic approach is adopted by assuming a less efficient heating mechanism, the solar wind problem can be approximated by a polytropic equation relating the pressure $P$ to the density through a power law of the following form
\begin{equation}
P = K \, \rho^{\gamma}  \, , 
\label{poleq}
\end{equation}
where $K$ is the polytropic constant and $\gamma$ is the polytropic index. 

 A special case is the adiabatic atmosphere with a polytropic index $\gamma=5/3$. In this case, a fluid element does not exchange heat as it propagates away from the surface of the Sun. A softly heated atmosphere has $ 1.1 < \gamma < 5/3$, the limited heated atmosphere corresponds to $\gamma =  1.1$, and the intensively heated atmosphere corresponds to $1.0 < \gamma < 1.1$ \citep{P1965rev}. Recent measurements suggest an effective polytropic index for the solar corona corresponds to $\gamma = 1.1 \pm 0.02$ \citep{VD2011}.

The EOS of an ideal gas (\ref{poleq})
\begin{equation}
P = \frac{\kappa}{\bar{m}} \, \rho \, T  \, ,
\label{poleqT}
\end{equation}
where $\bar{m}$ is the average molecular mass of the fluid with a typical value in case of the solar wind $0.6 \, m_p$ \citep{Mercier2015}, which is the one we use in this study for polytropic model. This allows us to express the sound speed at the base of solar wind $v_{s,b}$ for a polytropic gas of index $\gamma$ in the following form
\begin{equation}
v^2_{s,b} = \frac{dP}{d\rho} \mathrel{\bigg|}_{\rho=\rho_b}= \gamma \, K \, \rho_b^{\gamma-1} = \frac{\gamma \, \kappa T_b}{\bar{m}}  \, ,
\label{v_sb}
\end{equation}
where index $b$ refers to quantities at the base of the solar wind. As in the isothermal model, the base of the solar wind is located at the solar radius $R_{\odot}$ and the order of magnitude of the density there is $\rho_b = 10^8\bar{m}~$cm$^{-3}$ \citep{Newkirk1967, Mercier2015}.

Similarly to the isothermal case, we define dimensionless quantities for the radius $\xi=r/R_{\odot}$, density $\bar{\rho}(\xi)=\rho(\xi)/\rho_b$ and the Mach number 
\begin{equation}
M(\xi)=v(\xi)/v_s(\xi)=M_o(\xi)/\sqrt{\bar{\rho}^{\gamma-1}} \, ,
\label{machnum}
\end{equation}
where $M_o(\xi)= v(\xi)/v_{s,b}$ is the ratio of the flow velocity over the sound speed at the base of the corona ($R_{\odot}$). Since the temperature is not constant the sound speed is a function of radius as well, thus $v_s=v_s(\xi)$. 

Substituting the dimensionless expressions for $r, \, \rho$, $v$ into equation~(\ref{machnum}), and $dP/d\rho$ by equation~(\ref{v_sb}) in the conservation laws of mass equation~(\ref{syseq1}) and momentum equation~(\ref{syseq2}), we deduce a system of two ordinary differential equations that describes the polytropic model of the solar wind
\begin{equation}
\frac{d}{d\xi} \left( \bar{\rho} \, M_o \, \xi^2 \right) = 0 \, ,
\label{syseqpol1}
\end{equation}
and
\begin{equation}
M_o \frac{dM_o}{d\xi}\, + \frac{\lambda}{\xi^2} + \frac{1}{\gamma-1} \, \frac{d}{d\xi}\bar{\rho}^{\gamma-1} = 0\, .
\label{syseqpol2}
\end{equation}
where we introduce the dimensionless parameter $\lambda = \frac{G \, M_{\odot}}{r_b \, v^2_{s,b}}$ and $\mu=\bar{\rho} \, M_o \, \xi^2=\frac{\dot{M}}{4 \, \pi \, \bar{\rho}_b \, v_{s,b} \, r_b^2}$ is the dimensionless mass--loss rate or flow parameter and $\dot{M}$ is the mass loss rate. The last parameter can be evaluated by the following expression
\begin{equation}
\mu=\frac{v_b}{v_{s,b}} = \frac{\lambda^2}{2} \left[ \frac{2-2 \, \lambda \, (\gamma-1)}{5-3 \, \gamma} \right]^{\frac{1}{\gamma-1}-\frac{3}{2}}\, .
\label{eqmiu}
\end{equation}
where we have integrated the Bernoulli equation from base up to the critical point. Higher order terms of the ration $v_b/v_{s,b}\ll 1$ vanish and then we obtain the expression by solving the remaining equation for $v_b/v_{s,b}$ which is an other form of mass--loss rate. In addition, $\mu$ is related to the density function $\bar{\rho}(\xi)$ via equation
\begin{equation}
\bar{\rho} = \frac{\mu}{M_o \, \xi^2} \mathrel{\bigg|}_{\xi=\xi_c, \, M_o=M_c} = \left( \frac{4 \, \mu}{\lambda^2} \right)^{\frac{2}{5-3\, \gamma}}\, .
\label{eqrho}
\end{equation}
which is the following equation~(\ref{eqmiu}) combined with the expression~(\ref{critm}) for $M_c$. The value $\mu = \lambda^2/4$ corresponds to $\bar{\rho}_c = 1.0$ and the solution is identical to the one of isothermal model ($\gamma=1$).

The system of equations~(\ref{syseqpol1})--(\ref{syseqpol2}) leads to the dimensionless Bernoulli integral
\begin{equation}
\frac{M_o^2}{2} - \frac{\lambda}{\xi} + \frac{1}{\gamma-1} \, \left( \frac{\mu}{M_o \, \xi^2} \right)^{\gamma-1} = E\, .
\label{bereqpol}
\end{equation}
where $E$ is the dimensionless specific energy of the flow. The pair of parameters $(\mu, E)$ determines the flow and $M_o$ is a function of $(\xi, \mu, E)$.

Because of the existence of the critical point one needs to integrate the ordinary differential equation numerically. 

The minimization of the first derivative in $\xi$ of equation~(\ref{bereqpol}) gives the equation of the Mach number in the polytropic model (cf. equation~(3) in \citep{Habbal1983}, and equation~(2.2a) in \citep{Bailyn1985}, with $f=1, \, d=0$.) 
\begin{equation}
\frac{1}{M_o^2} \frac{dM_o^2}{d\xi}  = \frac{2}{\xi} \frac{2-\frac{\lambda}{\xi}\left( \frac{1}{\bar{\rho}^{\gamma-1}} \right)}{\frac{M_o^2}{\bar{\rho}^{\gamma-1}}-1} \, .
\label{macheqpol1}
\end{equation}

Then by substituting equation~(\ref{machnum}) into equation~(\ref{macheqpol1}) we obtain the Mach equation for the polytropic model
\begin{equation}
\frac{dM_o}{d\xi} = \frac{M_o}{\xi} \frac{2-\frac{\lambda}{\xi} \frac{1}{\bar{\rho}^{\gamma-1}}}{\frac{M_o^2}{\bar{\rho}^{\gamma-1}}-1} \, .
\label{macheqpol}
\end{equation}
for $\gamma=1$ we recover the isothermal wind equation~(\ref{veleq}). 
The critical point corresponds to $M=1$ and $\xi= \xi_c =\frac{\lambda}{2 M_o^2(\xi_c)}$ where $M_c=M_o(\xi_c)$. We refer to this in \S~\ref{paths}. In the last equation~(\ref{macheqpol}), we solve directly the system of equations~(\ref{syseqpol1}--\ref{syseqpol2}). Therefore, the solution of equation~(\ref{macheqpol}) will provide the expression of a polytropic wind.

\subsection{Mathematical singularities of the problem}
\label{paths}

The solution of the non--linear ordinary differential equation~(\ref{veleq}) by a direct numerical integration requires special treatment of the singularities of the problem. In particular, the critical point
\begin{equation}
\left( M_c=1, \qquad \, \xi_c=\lambda/2 \right) \, ,
\label{critp}
\end{equation}
in the isothermal wind ordinary differential equation~(\ref{veleq}), leads to an expression where the derivative $\frac{dM}{d\xi}$ becomes equal to an undefined expression $\left( \frac{0}{0} \right)$. This case can be resolved using the l'Hospital rule.

Application of the CPS leads to the removal the previous mathematical singularity and the solution of equation~(\ref{veleq}) by a direct numerical integration and without the aid of the Bernoulli integral through the technique of integration the right and the left side of (\ref{bereq}). Furthermore, CPS permits the calculation of the slope $\frac{dM}{d\xi}$ at each point of function $M(\xi)$ without application of l'Hospital rule.

In the less trivial case of the polytropic non--linear equation~(\ref{macheqpol}), a similar critical point occurs $( M_c,\, \xi_c)$. The evaluation of the parameters of the critical point is not as straightforward as in the isothermal wind, and its location is a function of the physical quantities of the polytropic fluid, $\xi_c(\lambda, \mu, \gamma)$ or equivalently of the density at critical point $\bar{\rho_c}$ which is given by equation~(\ref{eqrho} (figures~\ref{figisopolden}, \ref{figisopolgam})
\begin{equation}
\left( M_c = \sqrt{\bar{\rho}^{\gamma-1}}, \qquad \, \xi_c = \frac{\lambda}{2} \, \frac{1}{\bar{\rho}^{\gamma-1}} \right) \, .
\label{critm}
\end{equation}

The critical point $M_c=1$, $\xi=\xi_c$ can be calculated as a function of parameters $\lambda, \, \mu$ as  
\begin{equation}
\left( M_c = \left(\frac{4 \, \mu}{\lambda^2}\right)^\frac{\gamma-1}{5-3\, \gamma}, \quad \, \xi_c = \left(\frac{\lambda}{2}\right)^{\gamma+1} \left(\frac{1}{\mu^{2\, \gamma-2}}\right)^\frac{1}{5-3\, \gamma} \right) \, .
\label{critm2}
\end{equation}

\subsection{Numerical integration on the complex plane}
\label{integration}

CPS is a suitable numerical method for the  integration of ordinary differential equations on the complex plane, either along a real interval when the independent variable $r$ is real, or along a complex contour when $r$ is complex, to detour critical points. The key idea is to transform the real--valued functions of real variable, the ``real distance’’, $r \in \mathbb{R}$, into complex--valued functions of the ``complex distance’’, $r \in \mathbb{C}$, which constitutes the complex path along which integration proceeds, \citep{GV1994, GK2014}. A similar implementation has been developed in problems of dynamics where the time plays the role of the independent variable ``complex time’’ (e.g. \cite{Orendt2009}). This way, any indeterminate forms such as  $\left(\frac{0}{0}\right)$ appearing in the equations are removed. Provided that the imaginary part of the physical quantities remains small, one can ensure that the numerical solution approaches the actual solution of the problem.

We are interested in the real--valued function of one real variable $M(\xi)$, ($\xi$ is the ``real distance’’) to detour the critical points discussed in section \ref{paths}.
The initial conditions for equation~(\ref{veleq}) are 
\begin{equation}
M(\xi_i) =1,
\label{incon}
\end{equation}
at $\xi_i=\lambda/2$. 

The initial conditions of equation~(\ref{macheqpol}) are 
\begin{equation}
M_o(\xi_i) = M_c,
\label{incon}
\end{equation}
at $\xi_i=\xi_c$, where $M_c$ is given by equation~(\ref{critm}). 

Here we note that index $i$ indicates the value at the starting point of integration where the initial conditions are applied. 

In the framework of CPS, we introduce the function $M(\xi)=\bar{M}(\xi)+i\, \breve{M}(\xi)$ as a complex function of a complex variable $\xi=\bar{\xi}+i\,\breve{\xi}$. The integration contour $\xi$ corresponds to a straight--line segment parallel to the real axis and at a small imaginary $i\,\breve{\xi}$ from it. As a result initial conditions~(\ref{incon}) become

\begin{equation}
M_i = \bar{M}_i + i\, \breve{M}_i \, ,
\label{icon1}
\end{equation}
%
%
%
at $\xi_i=\bar{\xi}_i+i\,\breve{\xi}_i$, where the imaginary part $\breve{M}_i$ is zero or small compared to $\bar{M}_i$, respectively. More specifically, values of real and imaginary parts of each quantity at $\xi_i$ are 
\begin{subequations}
\begin{equation}
\bar{M}_i =1 \, , \qquad \breve{M}_i=10^{-4} \,  
\label{icon2a}
\end{equation}
\begin{equation}
\bar{\xi}_i =\frac{\lambda}{2} \, , \qquad \breve{\xi}_i=10^{-4} \, . \label{icon2c}
\end{equation}
\end{subequations}

In polytropic models, we introduce the complex function $M_o(\xi)=\bar{M}_o(\xi)+i\, \breve{M}_o(\xi)$ in the differential equation~(\ref{macheqpol}). The values of the real parts of the initial conditions $\bar{M}_o (\xi_i)$ and $\bar{\xi}_i$ are given by equations~(\ref{critm2}). All the other values of the initial conditions remain as the corresponding ones in equations~(\ref{icon2a}, \ref{icon2c}), respectively.

In the differential equation~(\ref{macheqpol}), we define the function $\bar{\rho}(\xi)=\mathrm{Re}(\bar{\rho}(\xi))+i\,\mathrm{Im}(\bar{\rho}(\xi))$ as a complex function of complex variable $\xi$. At the critical point~(\ref{critm2}, the value of $\mathrm{Re}(\bar{\rho}(\xi_i))$ is given by equation~(\ref{eqrho}) and the imaginary part is set to $\mathrm{Im}(\bar{\rho}(\xi_i))=0$. After the numerical integration we keep the real part $\mathrm{Re}(\bar{\rho}(\xi))$ which is used with $\bar{M}_o(\xi)$ in equation~(\ref{machnum}) to give the solution of Mach function $\bar{M}(\xi)$.

We integrate equation~\ref{veleq} on two complex contours, one from point $\xi_{start}=\xi_i$ forward to $\xi_{end1}=100 \times \lambda$ and an other backwards to $\xi_{end2}=0$. 

The accuracy of the solutions is related to choice of the imaginary part in the CPS. The estimation of the accuracy of this method is achieved through the solution of the equations a closed complex contour (initial and final point being the same). The integration error is equal to the difference of the solution at the end point minus the one at the start point, which formally should have been equal to zero \citep{GV2012}.

\begin{figure}
\includegraphics[width=0.5\textwidth]{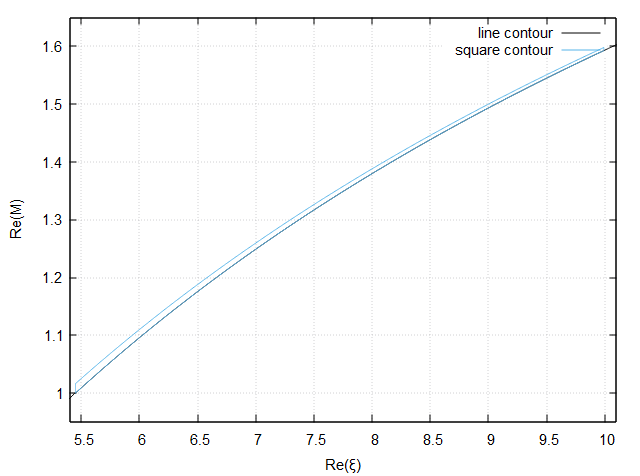}
\quad
\includegraphics[width=0.5\textwidth]{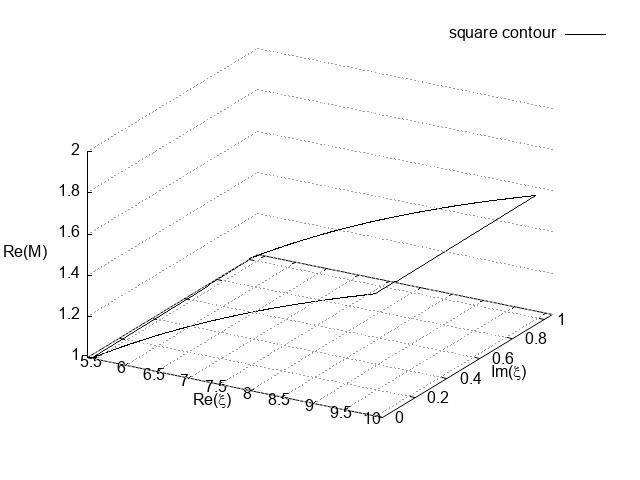}
\caption{\label{square-zoom-2} Solution of equation~(\ref{veleq}), $\bar{M}$ vs $\bar{\xi}$, under initial conditions $\bar{M}_i=1$ and $\breve{M}_i=10^{-4}$ via the linear complex contour~(\ref{icon2c}) and a square complex contour (upper figure). Solution of equation~(\ref{veleq}), $\bar{M}$ vs integrating square contour in complex plane $\xi_1 \, (\bar{\xi}_i, 0)$, $\xi_2 \, (10, 0)$, $\xi_3 \, (10, 1)$, $\xi_4 \, (\bar{\xi}_i, 1)$, where $\bar{\xi}_i=5.45$ (lower figure).}
\end{figure}

We then focus on the implementation of CPS in this problem by integrating equation~(\ref{veleq}) in different two complex contours. The ``direct" one that is described in the previous paragraph and one square contour as defined by the edges $\xi_1 \, (\bar{\xi}_i, 0)$, $\xi_2 \, (10, 0)$, $\xi_3 \, (10, 1)$, $\xi_4 \, (\bar{\xi}_i, 1)$ where the first value in the coordinates is the real value $\bar{\xi}$ and the second is the imaginary value $\breve{\xi}$. The integration step is $10^{-2}$ both along the real and imaginary axis and the initial conditions are the same as in equations~\ref{icon2a}, \ref{icon2c}. The real parts of the numerical solutions in both cases are identical as we can see in figure~\ref{square-zoom-2}. In the upper panel, we show the deviation between the square contour and the line contour. The lower panel refers both to the closed complex contour $\xi$ that we use to integrate the ordinary differential equation and to the real part of the Mach number function, $\bar{M}$ we obtain from the application of CPS. 

The physical meaning of the physical quantity is encapsulated in the real part of the solution $\breve{M}$, whereas, the imaginary part $\breve{M}_i$ is an auxiliary quantity allowing the integration of the equation and needs to remain small, similar to its initial value. Therefore, the values of the imaginary parts of the functions involved in our problem $\breve{M}$, $\breve{M}_o$, and $\mathrm{Im}(\bar{\rho}(\xi))$ converge to sufficiently small quantities at the same or smaller order of magnitude with the initial one. 

In general, because of the requirement for the real part of the solution $\bar{M}$ to remain in a physical framework, conceding the fact that we already know the values of the solution, we approach empirically the appropriate initial value of the imaginary parts of the functions ($\breve{M}$, $\breve{M}_o$ etc), such as the value of $\breve{M}$ is much smaller than the value of $\bar{M}$ and the latter remains physically accepted. This approximation is based on the main aim of CPS to remove the singularity at the critical point by assuming a small imaginary part at integrating contour, $\breve{\xi}_i$, as well as a small one (at the same point) of the imaginary part of a complex function in comparison to its real part. Larger imaginary parts omay lead to other solutions without physical meaning. For example, they can be related to a discontinuity at the critical point. 

In the present problem the value of $\breve{M}$ acts as a parameter which for positive values leads to the Parker solution and for negative values to the Bondi one. This issue is discussed in \S~\ref{Isores}

At this point we need to clarify some properties related to the CPS. The CPS does not calculate the integral of a complex function along a closed contour in the complex plane (such as equations~(\ref{syseqpol2}, \ref{bereqpol}), but integrates numerically an ODE of a complex function (such as equations~(\ref{veleq}, \ref{macheqpol}) on the complex plane. This problem does not contain poles in the region of interest but only one singularity which is used as initial point of the numerical integration, as we have discussed about this issue in \S~\ref{paths}. This is the reason that CPS is an appropriate method for solving this problem. CPS under some appropriate modifications could be applied in cases of ordinary differential equations contain logarithms or power laws forms as it is presented in detail in \cite{GK2014}.

\section{Results}
\label{results}

The code for our simulations is written in Fortran and for compiling we use the GNU Fortran compiler, gfortran, which belongs to the GNU Compiler Collection (http://gcc.gn u.org/) and is licensed under the GNU General Public License (http://www.gnu.org/licenses/gpl.html). It has been installed by the TDM-GCC ‘‘Compiler Suite for Windows’’ (http://tdm-gcc.tdragon.net/), which is free software distribu\-ted under the terms of the GPL. We use the Fortran package dcrkf54.f95 \citep{GV2012}, a Runge-–Kutta-–Fehlberg code of fourth and fifth order, appropriately modified for the solution of complex initial value problems, with highly complex expressions for their ordinary differential equations along contours (not necessarily simple or closed) prescribed as continuous chains of straight--line segments. 

In what follows, we refer and plot only the real part of the solution $\bar{M}(\xi)$ and $\xi$ at figures refers to $\bar{\xi}$. 

\subsection{Isothermal Wind results}
\label{Isores}
We integrate Mach equation~(\ref{veleq}) assuming the initial conditions in equations~(\ref{icon2a}-\ref{icon2c}). The integration is performed on a complex contour with imaginary part $\breve{\xi}_i=10^{-4}$ and $\breve{M}_i=10^{-4}$. We obtain the Parker isothermal wind solution, as the one resulting from the Bernoulli equation~(\ref{bereq}) for $M_c=1$ and $B=B_c$ which is the physically acceptable solution. We set the temperature at the base of the corona $T_b$, which is used in the solution of equations~(\ref{v_so1}) and (\ref{eqlam}). We note that a choice of a negative initial imaginary part i.e.~$\breve{M}_i=-1 \times 10^{-4}$ leads to Bondi accretion solution.

Integration of the Mach equation (\ref{veleq}) with initial conditions equal to $\bar{M}_i=1.5, \, 2.0$ and $\breve{M}_i=10^{-4}$ via a complex contour (\ref{icon2c}) leads to the same solution of the Bernoulli equation (\ref{bereq}) for $M_c>1$ and $B<B_c$, which corresponds to a supersonic solar wind velocity everywhere with a local minimum at the vicinity of the critical point.

We further integrate the Mach equation~(\ref{veleq} assuming the following initial conditions $\bar{M}_i=2 \times 10^{-1}, \,5 \times 10^{-1}$ and $\breve{M}_i=10^{-4}$ on the same complex contour~(\ref{icon2c}). We obtain the solution of the Bernoulli equation~(\ref{bereq}) for $M_c<1$ and $B<B_c$ which corresponds to a breeze, a flow that is everywhere subsonic, with a local maximum at at the vicinity of the critical point. While this solution is mathematically valid, it cannot represent the flow of the solar wind as it never becomes supersonic. 
 
\begin{figure}
\includegraphics[width=1.0\textwidth]{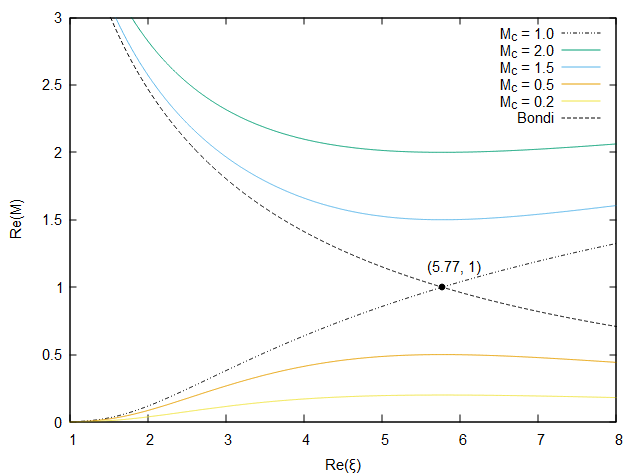}
\caption{\label{fig:figall} The solutions obtained with the CPS of $\bar{M}$ versus $\bar{\xi}$, of all cases for an isothermal spherical flow.}
\end{figure}

The above solutions are plotted in figure~(\ref{fig:figall}) where the temperature of the isothermal wind is set to $T_b = 10^6$K and the critical point is located at $\bar{\xi}_i=5.77$. 

As shown in figure~(\ref{fig:figall}), we can distinguish unphysical solutions from the physical ones. The Bondi accretion solution and the ones noted as $M_c > 1$ have a super--sonic flow at the base of the corona. This behaviour is not consistent with a static solar photosphere. Solutions with $M_c \leq 1$ correspond to a sub--sonic flow in the vicinity of $\xi_c$ and cannot describe the solar corona \citep{Fitz2014}. 

We investigate the role of the complex contour in the integration of Mach equation~(\ref{veleq} by repeating the previous computations via a real contour, where $\breve{\xi}=0$. For subsonic $(\bar{M}_i < 1)$ or supersonic solutions $(\bar{M}_i > 1)$ we recover the same solutions as in 
figure~(\ref{fig:figall}). If we use the  initial conditions of equation~(\ref{icon2a}) the profile of velocity changes and we take an unphysical solution, with a discontinuity in the derivative of $M(\xi)$, instead of the expected smooth solution.

\begin{figure}
\includegraphics[width=0.5\textwidth]{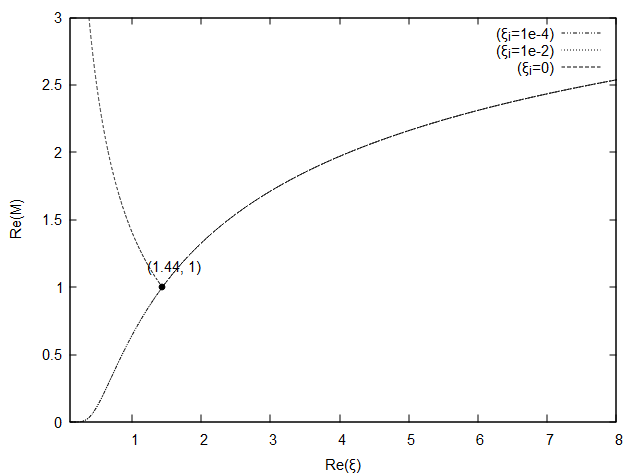}
\quad
\includegraphics[width=0.5\textwidth]{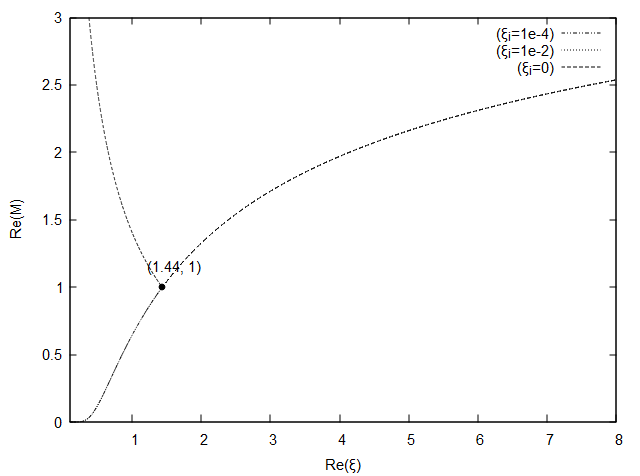}
\caption{\label{fig4uni}Solution of equation~(\ref{veleq}), $\bar{M}$ vs $\bar{\xi}$ in two cases for $\breve{\xi}=10^{-2}$ and $\breve{\xi}=10^{-4}$, under initial conditions with real part $\bar{M}_i=1$ and two different values of imaginary part, $\breve{M}_i=10^{-4}$ (upper panel) and $\breve{M}_i=10^{-2}$ (lower panel) via a real contour, where $\breve{\xi}=0$.}
\end{figure}

The solutions for $\bar{M}(\xi)$, for the integration of the complex function $M(\xi)$ on the real and complex contours $\bar{\xi}$, $\xi$ respectively, is shown in figure~(\ref{fig4uni}). The upper panel contains the solutions of equation~(\ref{veleq}) for integration on a complex contour with imaginary part $\breve{\xi}=10^{-2},\, 10^{-4}, \, 0$ where $\bar{M}_i=1$ and $\breve{M}_i=10^{-4}$ for $T_b = 4 \times 10^6$K. The lower panel shows to solutions of equation~(\ref{veleq}) for integration in the same complex contour but for initial conditions where the real part remains the same $\bar{M}_i=1$ but the imaginary is set to $\breve{M}_i=10^{-2}$. In all cases the solutions $\breve{\xi}=10^{-2},\, 10^{-4}$ are identical to each other. In addition, the behaviour of solution for $\breve{\xi} > 0$ does not change even if $\breve{\xi} < 0$ so their plots are omitted in figures~(\ref{fig4uni}). The form of the solution changes only for the integration on real contour. This result confirms the validity of CPS as an appropriate method which providing the physical solutions of the solar wind problem normally by avoiding numerical problems and without restrictions in values of its parameters. We note that there is the restriction of $\bar{M}_i \neq 0$ because of the singularity which appears in denominator of equation~(\ref{veleq}) at the critical point. Also, we examine the case of integration on a complex contour $\breve{\xi}=10^{-4}$ with critical point $\bar{M}_i=1$ and $\breve{M}_i=-10^{-2}, \, -10^{-4}$ for $T_b = 4 \times 10^6$K and recover the Bondi accretion solution in both cases.

\begin{figure}
\includegraphics[width=1.0\textwidth]{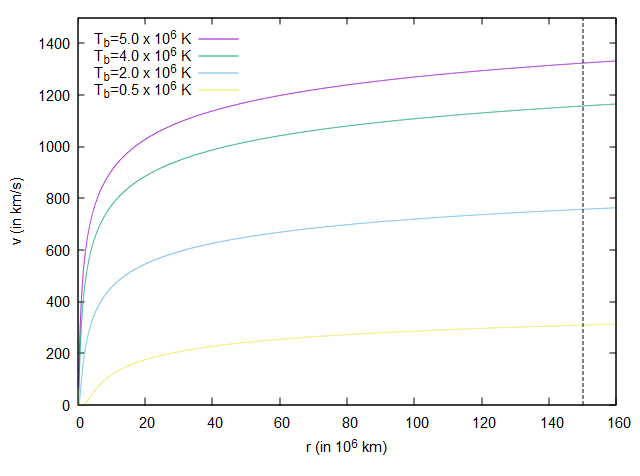}
\caption{\label{figvsvr} Solar wind velocity $v$ versus distance $r$ form the solar corona for various temperatures $T_b$. The dashed line is the orbit of Earth which is at $150 \times 10^6$ km. 
Here we have to mention that the centre of the sun corresponds to the origin of axis in all other figures of this study.} 
\end{figure}

In figure~\ref{figvsvr}, we plot the CPS solution for the Parker solar wind, where $v = M \, v_s$ and $r$ is the distance from the sun. We verify that for temperatures in the physically accepted range, $T\sim 1$-$2\times 10^6$K, the flow velocity is in the range of  hundreds of kilometers per second at 1 AU. The critical surface where the solar wind makes the transition from subsonic to supersonic flow lies a few solar radii away from the Sun (i.e., $r_c\sim 5\,R_\odot$) \citep{Priest1987, Fitz2014}.

\subsection{Polytropic winds}

Since we have been able to recover the Parker isothermal wind solution, we further investigate the more complicated solutions of polytropic models. We integrate equation~(\ref{macheqpol}) with initial conditions~(\ref{critm2} for the real parts of equations~(\ref{icon2a}) and (\ref{icon2c}). Equation~(\ref{macheqpol}) is an ordinary differential equation of the $M_o$, the flow velocity scaled to the speed of sound at the base of the solar corona, therefore, we use equation~(\ref{machnum} so that we recover the Mach number $M$ at the end of numerical integration. The initial condition are the ones mentioned above correspond to $\bar{M_o}$.

We have made three groups of numerical runs so that we investigate the behaviour of the numerical solution subject to the following three parameters: the polytropic index, the flow parameter (mass--loss rate) and the temperature. 

First, we illustrate the role of the polytropic index $\gamma$ and the flow parameter $\mu$ at the critical point $\bar{\rho}_c$ in the velocity. Here, we set the above quantities as well as the temperature at the coronal base, $T_b$. These quantities are needed for the evaluation of $\lambda$ and the sound speed at the coronal base, $v_{s,b}$. We also explore the solutions for different mass-loss rate parameters. Then, we construct the solution for solar wind by providing only the values of ($\gamma, \, T_b$) and after that, we calculate the values of $\mu$, $\bar{\rho}_c$ and $\lambda$. All these computations are done with an integration step $10^{-3}$ both  forward and backward. The imaginary parts have values $\breve{\xi}_i=10^{-4}$ and $\breve{M}_i=10^{-4}$.

\subsubsection{Polytropic index dependence}

\begin{figure}
\includegraphics[width=1.0\textwidth]{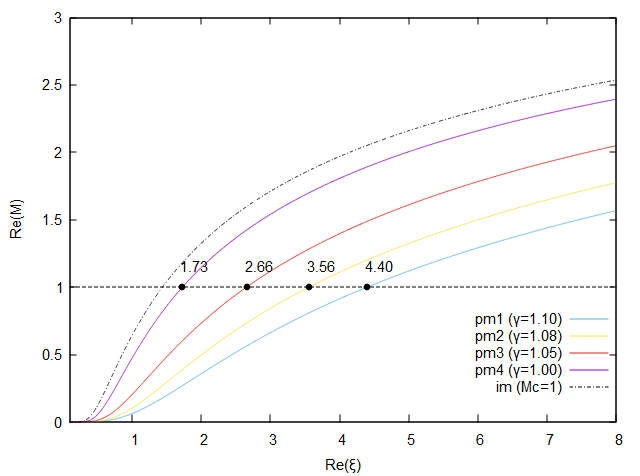}
\caption{\label{figisopolgam} Solutions of $\bar{M}$ versus $\bar{\xi}$ for isothermal model with $\bar{M}_i=1$ in comparison to four different cases of polytropic model with $\gamma = 1.00, \, 1.05, \, 1.08, \, 1.10$. All results are with $T_b=4\times10^6$ K and $\mu= 4 \times 10^{-4}$.}
\end{figure}

In figure~(\ref{figisopolgam}), we plot the solution of $\bar{M}$ versus $\xi$ for the isothermal model with $\bar{M}=1$ in comparison to four cases of polytropic indices $\gamma = 1.00, \, 1.05, \, 1.08\, 1.10$. All results are obtained for given temperatures and mass loss at the surface of the Sun $T_b=4\times10^6$K and $\mu= 4 \times 10^{-4}$. We confirm that in each polytropic model the location of the critical point $\left( \bar{M}_i, \, \bar{\xi}_i \right)$, changes for the different polytropic index. 

One important consequence of the application of CPS in the polytropic model of solar wind is that we find the physical solutions for every value of $\gamma$ in $\xi \rightarrow 0$ for $M_o \rightarrow 0$, and from computational and mathematical point of view, we can extend our calculation in a range of $\gamma > 3/2$, although in this region there are no wind solutions with physical meaning. As mentioned in \cite{Bailyn1985} above this limit the solution becomes supersonic for $\xi \rightarrow 0$ and $M_o \rightarrow \infty$ which is a nonphysical behaviour and should be rejected.

We compare the solutions for the Parker isothermal model against five polytropic models with different $\gamma$ in figure~(\ref{figisopolgam}) where we keep the same flow parameter $\mu= 4 \times 10^{-4}$, and $T_b$ is given for each case from equation~(\ref{eqmiu}). We note that in each case the density at the critical point varies according to equation~(\ref{eqrho}) for the same value of parameter $\mu$. The deviation between the solutions for isothermal model, note in figure as (im), and the corresponding one for $\gamma = 1.00$, note in figure as (pm4), is because of the different assumption in the molecular mass of fluid. 

In figure~(\ref{figisopolgam}) it is shown that as the polytropic index increases, the critical point shifts to larger distance from the star and the transition from subsonic to supersonic flow becomes smoother laeding to lower velocities at large distances.  

\subsubsection{Flow parameter dependence}

\begin{figure}
\includegraphics[width=1.0\textwidth]{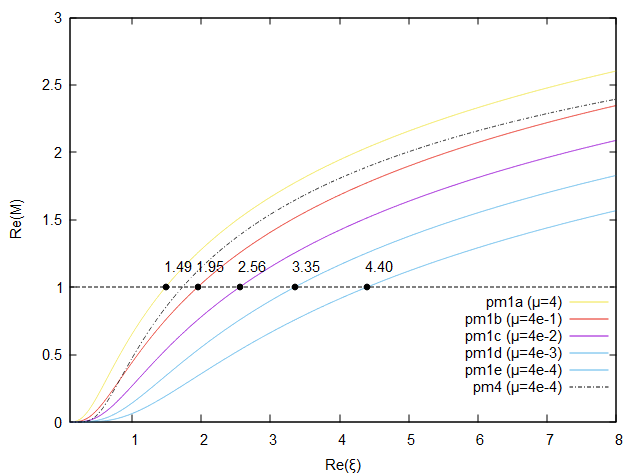}
\caption{\label{figisopolden} Solutions of $\bar{M}$ versus $\bar{\xi}$ for five cases of polytropic model with $\gamma = 1.10$ and $\mu = 4 \times 10^{-4}, \, 4 \times 10^{-3}, \, 0.04, \, 0.4, \ 4.0$, respectively. All results are for $T_b=4\times 10^6$K. }
\end{figure}

We have also studied the role of the flow parameter $\mu$ or equivalently of the density $\bar{\rho}_c$ in the polytropic model by solving for model pm1 ($\gamma = 1.10$) assuming six different values of flow parameter $\mu = 4\times 10^{0}, \,10^{-1}, \, 10^{-2}, \, 10^{-3}, \,10^{-4}$ and compare against pm4 ($\gamma = 1.00$, $\mu = 4\times 10^{-4}$). We plot the solutions in figure~(\ref{figisopolden}). 

In both figures~\ref{figisopolgam} and \ref{figisopolden}, the intersection of the horizontal line for $M=1$ with each curve are in a different point so the critical point is changing in each case. The variation of critical point according to the value of $\gamma$ are discussed in \S~4.1 of \cite{Keppens1999} where their different positions are indicated in figure~(1) there.

Also, as the flow parameter decreases, while keeping the same polytropic index, the transition point shifts further from the star and the wind transitions from subsonic to supersonic with a smaller derivative. This leads to lower velocities. Following the opposite route and starting by assuming the profile pm1e (identical to pm1 shown in figure~\ref{figisopolgam}) with high $\mu$ to pm1a, we notice that the increase of $\mu$ leads to  higher values of velocities. As a result, profiles of pm1b and pm1a intersect the profile of pm4.

\subsubsection{Temperature dependence}

\begin{figure}
\includegraphics[width=1.0\textwidth]{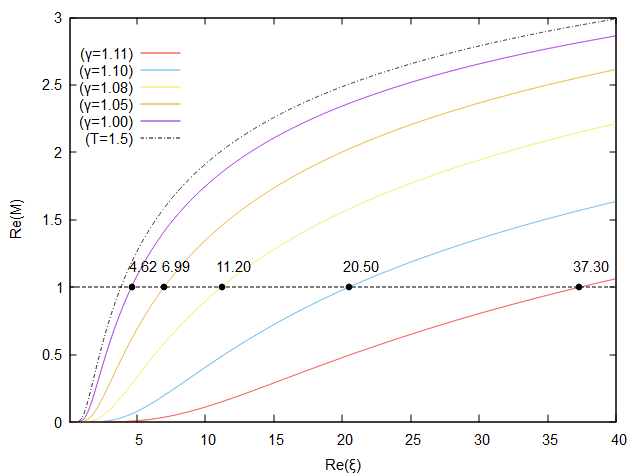}
\caption{\label{T15vargamma} Solutions of $\bar{M}$ versus $\bar{\xi}$ in comparison for five cases of polytropic model with $\gamma = 1.11, \, 1.10, \, 1.08, \, \,1.05, \,1.0$ and the corresponding isothermal model in same temperature. All results are with $T_b=1.5\times10^6$ K and have $\mu= 2 \times 10^{-1}$, the first 3 models and $\mu= 1.5$ the last one. Also, $\bar{\rho}_c = 5.1 \times 10^{-2}, \,5.5 \times 10^{-2}, \, 6.0 \times 10^{-2}, \,0.5$, respectively.}
\end{figure}

\begin{figure}
\includegraphics[width=1.0\textwidth]{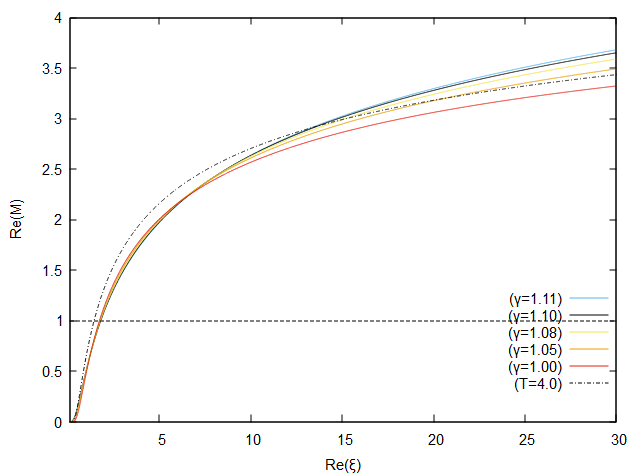}
\caption{\label{T4vargamma} Solutions of $\bar{M}$ versus $\bar{\xi}$ in comparison for five cases of polytropic model with $\gamma = 1.11, \, 1.10, \, 1.08, \, 1.05, \,1.0$ and the corresponding isothermal model in same temperature. All results are with $T_b=4\times10^6$ K and have $\mu= 6.2 \times 10^{-6}, \,1.2 \times 10^{-4}, \, 9.1 \times 10^{-4}, \,  10.7$, respectively. Also, $\bar{\rho}_c = 2.5 \times 10^{-8}, \,1.25 \times 10^{-6}, \, 2.1 \times 10^{-5}, \,0.5$, respectively. The x-coordinates of their critical points are $\bar{\xi}_i=1.81, \, 1.80, \, 1.78, \, 1.76, \, 1.73$, and $ 1.44$ for the isothermal model.}
\end{figure}

We have also performed a second group of numerical experiments to study the role of temperature at the coronal base, $T_b$ and polytropic index, $\gamma$, in polytropic models. We plot models with $\gamma = 1.10, \, 1.08, \, \,1.05, \,1.0$ for two different temperatures $T_b=1.5\times10^6, \, 4\times10^6$K in figures~(\ref{T15vargamma}, \ref{T4vargamma}). We further plot the isothermal models in the same temperatures for $\bar{M}_i=1$ and $\breve{M}_i=10^{-4}$ which are marked in these figures as $T_b=1.5$ and $4$, respectively, for comparison. These computations have been made with integration step $10^{-3}$ both forward and backward from the critical point. The imaginary parts has values $\breve{\xi}_i=10^{-4}$ and $\breve{M}_i=10^{-4}$ and the initial conditions takes the values calculated by equation~(\ref{critm2}).

In figure~(\ref{T15vargamma}), we notice that for low $T_b$ the increase of $\gamma$ do not change the values of $\mu$ and $\bar{\rho}_c$ drastically but both are higher than the corresponding ones for models in figure~(\ref{T4vargamma}) where the values of $T_b$ is higher. Specifically the difference in $\mu$ is from 3 to 5 times of order and in $\bar{\rho}_c$ is from 3 to 6 times of order. This fact drives to a higher differences in values of initial conditions (see equation~(\ref{critm2})) between the models in figure~(\ref{T15vargamma}) than the difference in comparison to the models in figure~(\ref{T4vargamma}). In figure~(\ref{T15vargamma}), the differences of values of $\mu$ and $\bar{\rho}_c$ between models are low but their values as absolute magnitudes (cause of $T_b$) are high so the curves of them are very different. In figure~(\ref{T4vargamma}), values of $\mu$ and $\bar{\rho}_c$ between models are higher but their values as absolute magnitudes (because of $T_b$) are low so the curves of them are slightly different.

As $T_b$ increases $\mu$ increases as well, by examining the isothermal model or the same $\gamma$ in polytropic model. For example, the order of magnitude of $\mu$ is $10^{-1}$ in case of $T_b=4\times10^6$K and $10^{-5}$--$10^{-3}$ in case of $T_b=1.5\times10^6$K, according to (\ref{eqmiu}). From our numerical experiments we verify that for $T_b \geq 3\times10^6$K, the solutions cross each other because the value of $\mu$ is comparable to values of $\lambda$, $M_o$, and $\xi$ in equation~(\ref{macheqpol}). Consequently, the resulting differential equation is quite different in such cases. This is shown in figure~(\ref{T4vargamma}). The same effect is presented in figure~(\ref{figisopolden}) where the curve of pm4 intersects the others.

If we compare the solution of model $T_b=1.5\times10^6$K with the corresponding one for $\gamma = 1.0$ in figure~(\ref{T15vargamma}) and the one of $T_b= 4\times10^6$ with the curve for $\gamma = 1.0$ in  figure~(\ref{T4vargamma}), we note that they are slightly different although they refer to the same model. This difference owing to the different assumption in molecular mass of fluid. In isothermal models we assume $\bar{m}=m_p/2$ but in polytropic models we assume $\bar{m}=0.6m_p$ (see \S~\ref{polysetup}).

\section{Discussion}
\label{discussion}

The purpose of this study is to revisit the problem of the solar wind by solving it via the CPS, demonstrating that this numerical technique is applicable and useful in astrophysical problems with critical points, singularities, undefined or indeterminate forms in the integration contours of their differential equations. 

While the CPS is not widely used, it is indeed an appropriate numerical method that avoids mathematical pathologies that appear in initial value problems. Ordinary differential equations are integrated directly without using other algebraic techniques and one obtains the physically acceptable solutions. In particular, here the existence of a critical point does not allow, using conventional methods, the integration from this particular point, but rather one needs to start from a point just before or after. The implementation of CPS allow us to use exactly the critical point as initial point of numerical integration and then move forward and backward in order to take the physical solution, at the expense of using a complex valued function.    

If we try to solve equations~(\ref{macheq}) or (\ref{macheqpol}) by a simple ``shooting method'' using a common solver we have to approximate the initial condition $M_c=1$ at the critical point before and after it by the supersonic and subsonic solution respectively and then use interpolation at this point to obtain the physical solution. For instance, if we would like to get the physical solution for the isothermal model, we can solve equation~(\ref{macheq}) for two different initial conditions $M_c = 1^+, 1^-$ and in each trial we shift the initial condition closer to one, see how the the solutions approach the physical one as $M_c \rightarrow 1$ the figure~\ref{fig:figall}. This treatment needs multiple runs until fine tuning is achieved, plus it is needed an interpolation. On the other hand, by using l'Hospital rule we need an analytical solution which may complicate the calculation. 

In addition, the integration path has a singularity at the origin of the real axis ($\bar{\xi}=0$) as $\xi$ appears at the denominator in the ordinary differential equations~(\ref{macheq}) and (\ref{macheqpol}). This singularity vanishes if the integration proceeds on a complex contour parallel to the real axis on a distance equal to imaginary part $\breve{\xi}$. Furthermore, when we follow the complete CPS and transform the real function $M(\xi)$ to a complex one, we are able to integrate the ordinary differential equation~(\ref{macheq}) on a real axis by avoiding the singularity at the origin but we do not get the physical solutions, see figure~(\ref{fig4uni}). By using the CPS as described in \S~\ref{integration} we avoid smoothly all the mathematical pathologies of the problem and verify the topology of solutions both for isothermal and polytropic solar wind models.

Also, we check the Bernoulli integral both in isothermal and polytropic models by calculating $B$ in equation~(\ref{bereq}) and $E$ in equation~(\ref{bereqpol}) using the solutions obtained by CPS. Indeed, the numerical experiments show a deviation of these values within the numerical error of the numerical method.

In conclusion, the implementation of CPS in problem of solar wind velocity yields the physical solution of the problem directly through a numerical integration in cases of our study. This allows someone to test other promising models for describing solar wind flows from one hand plus to investigate the existence of other solutions in the topology of existing ones or further interesting physical characteristic, from the other. Overall, CPS is a promising numerical method for studying and other flow problems in astrophysics.

From the physical point of view, we verify the displacement of sonic point, for polytropic models as the the polytropic index increases. More specifically as the polytropic index increases the sonic point $\xi_c$ (for $M_c=1$) shifted to a further distance from the coronal of the star, as we shown in figures~(\ref{figisopolgam}, \ref{T15vargamma}) and (\ref{T4vargamma}). This is a consequence of the fact that a higher polytropic index leads to a less efficiently heated atmosphere around from the stars and at the values of $\gamma = 5/3$ it becomes adiabatic. Consequently, an isothermal atmosphere gives a critical point closer to the star and the adiabatic environment gives a critical point further out, keeping all other parameters the same. This behaviour is physically expected because an intensively heated atmosphere provides more energy to the wind which obtains a supersonic velocities earlier than in adiabatic case. However, in polytropic model is not allowed to examine cases up to $\gamma = 5/3$ because the solar parameters set restriction in polytropic index within a range of $[1, 1.183]$ \citep{Keppens1999, Wat2012}.

The observations of the acceleration region of the fast solar wind that have been done by UVCS shown that the wind could not be accelerated by thermodynamic expansion in case of the Sun \citep{Kohl:1998}. In addition, although Parker solar wind model predicts that the wind should make the transition to supersonic flow at about 5 solar radii far from the surface, the transition at sonic point appears to be located at only 1 solar radius above the photosphere. This fact suggests that there is some other mechanism, yet to be found, which accelerates the solar wind and is not predicted by theory of Parker, \citep{Priest1987, Fitz2014}. As a consequence, it is necessary either to improve the existing theory or to check other alternative theories of solar wind. In this direction CPS seems to be a useful technique for solving the problem of solar or star outflows and carrying out these tests. Of course, CPS does not provide an different physical explanation for the solar wind but only an alternate numerical method which can be applied in the existing solar wind theory.

Concerning the applicability of the CPS method to systems of ordinary differential equations, we need to mention that it has been tested in similar problems. A typical case is the study of magnetic field of a neutron star \citep{GK2011}, where the relativistic Grad--Shafranov equation is solving as a nonhomogeneous Sturm--Liouville problem with nonstandard boundary conditions. All the implementations of CPS concern problems so far, have been made on systems of ordinary differential equations. Therefore CPS constitutes a suitable method for studying several models of self--similar wind outflows or similar astrophysical problems.

\section{Conclusion}
\label{conclusion}

In the present investigation we have examined the one dimensional steady problem of the solar wind using the CPS demonstrating that this numerical method is appropriate for solving initial value problems of astrophysical flows, especially the simulations that assume a polytropic model in their framework. The ease of solutions, which are computationally less expensive compared to corresponding simulations of the time--dependent problem has allowed us to explore the parameter space focusing on the role of temperature, adiabatic index and mass loss rate in the solution. 


Some other challenging issues which are worth investigation are whether CPS can be generalised to relativistic Parker winds by using Taub equation of state \citep{Taub1948}, to magnetized astrophysical flows by using Weber--Davis mo\-del \citep{Weber1967} containing more than one critical point, to rotating winds \citep{Keppens1999}, to disc winds \citep{Wat2012}, or to self similar solutions which deal with non linear ODEs. All these subjects are beyond the purpose of the present paper and we intend to study them in subsequent bunch of publications.


\section{Acknowledgements}
VK and KNG thank Nektarios Vlahakis and Kanaris Tsin\-ganos for insightful discussions on the solutions of polytropic winds. The simulations where performed on RIGIL Computing System of the University of Patras, funded by the research programme of the University of Patras ELKE FK-80951.

\bibliographystyle{unsrtnat}
\bibliography{references}  






\end{document}